\documentclass[preprint,showpacs,aps,pra]{revtex4-1}
\usepackage{epsfig}
\usepackage{psfrag}
\usepackage{graphicx}
\usepackage{titlesec}
\usepackage{graphics}
\usepackage{amsmath}
\usepackage{color}
\usepackage[utf8]{inputenc}
\usepackage[title]{appendix}

\begin{document}

\title{Probing modified plasma waves in non-linear electrodynamics}

\author{Leonardo P. R. Ospedal} \email{leoopr@cbpf.br}
\affiliation{Centro Brasileiro de Pesquisas F\'isicas, Rua Dr. Xavier Sigaud
150, CEP 22290-180, Rio de Janeiro, RJ, Brazil}
\affiliation{Instituto de F\'{\i}sica, Universidade Federal do Rio Grande do Sul, Av. Bento Gon\c{c}alves 9500, CEP 91501-970, Porto Alegre, RS, Brazil}

\author{Fernando Haas} \email{fernando.haas@ufrgs.br}
\affiliation{Instituto de F\'{\i}sica, Universidade Federal do Rio Grande do Sul, Av. Bento Gon\c{c}alves 9500, CEP 91501-970, Porto Alegre, RS, Brazil}

\begin{abstract}

Properties of modified plasma waves in non-linear electrodynamics are investigated. We consider a cold, uniform, collisionless, and magnetized plasma model. Initially, we also assume small amplitude waves and the non-relativistic approximation. For electrostatic waves, we obtain a modi\-fied Trivelpiece–Gould dispersion relation with a suitable change in the plasma frequency and analyze the stability of modes. Furthermore, electromagnetic waves related to the generalized Appleton–Hartree equation are established. In this case, we discuss modifications in circularly polarized waves, ordinary and extraordinary modes. After that, we apply our results to particular cases of low-energy quantum electrodynamics and a generalized Born–Infeld model. The correspondent dispersion relations and effects on the propagation regions are determined. Finally, we include the relativistic and large amplitude effects for circularly polarized waves. We obtain the dispersion relation within effective non-linear electrodynamics and examine the behavior of the refractive index when the frequency of the propagating wave converges to the plasma frequency.

\end{abstract}

%

\maketitle

\section{Introduction}

In $1933$, based on the seminal works by Dirac \cite{Dirac_1928,Dirac_1930,Dirac_1931} about the quantum theory for the electron and the concept of quantum vacuum, Halpern pointed out that virtual electron-positron pairs could generate light-by-light scattering \cite{Halpern}. Subsequently, in $1934$, Heisenberg published two papers \cite{Heisenberg_1,Heisenberg_2}, where the connection between quantum vacuum fluctuations and light-by-light scattering was formulated in more detail. With these ideas and the initial development of quantum electrodynamics (QED), in $1935$, Euler and Kockel considered low-energy quantum effects and obtained non-linear corrections to Maxwell electrodyna\-mics. Interestingly enough, they also figured out the first calculation for the light-by-light cross section in the low-frequency regime \cite{Euler_Kockel}. Thereafter, in $1936$, this non-linear electrodynamics was generalized by Heisenberg and Euler \cite{Heisenberg_Euler}, where the authors included high-order quantum corrections. In parallel, from a classical viewpoint, Born and Infeld presented a non-linear model to avoid  the electric field singularity at small distances \cite{Born_Infeld}. Since then other non-linear electrodynamics have been proposed with different motivations, such as effective theories, proposals beyond the Standard Model of elementary particles, novel black-hole solutions, and applications in Cosmology (see, for instance, the reviews \cite{Plebanski,Sorokin_Notes} and references therein).


It is appropriate to highlight the renovated interest in non-linear electrodynamics due to the upgrade of some experiments. In 
$2017$, the ATLAS Collaboration carried out a first measurement of light-by-light scattering in heavy-ion collisions \cite{ATLAS}. After that, in $2018$, the CMS Collaboration also described similar measurements \cite{CMS}. More recently, in $2021$, the CMS and TOTEM Collaborations have reported a first search for light-by-light scattering in proton-proton collisions \cite{CMS_TOTEM}.  Bounds on cross sections have been used to test the Standard Model and QED predictions, as well as to constrain the new parameters associated with other proposals of non-linear electrodynamics. Along the same line, we indicate the work of ref. \cite{Ellis_2022} for some prospects at future colliders.


Furthermore, the development of high-intensity lasers  has also triggered a growing perspective to probe photon-photon and photon-plasma interactions \cite{Marklund_Lundin_EJPD,Piazza_RMP,Battesti_PhysRept,Karbstein,Fedotov,Ahmadiniaz,Marklund_Shukla_RMP}, 
 where the vacuum non-linearity and collective effects may induce interesting phenomena, such as
dispersive contributions to the wave propagation, electron-positron pair creation and new amplitude-dependent modes. For instance, an investigation of strong laser wave propagation in plasma was carried out in ref. \cite{Piazza_POP_2007}. By considering the Euler-Kockel electrodynamics, the authors obtained a non-trivial behaviour of the refractive index when the wave frequency approaches the effective plasma frequency. An experimental setup was also proposed, but the effect still remains to be detected.  Mention should be made that the astrophysical environments provide important situations involving strong magnetic fields, where the impact of non-linear electrodynamics and plasma effects  are also relevant. For example, ordinary neutron stars can reach magnetic fields of the order of $10^6-10^9 \, T$, while  magnetars can arrive at $10^{10}-10^{12} \, T$. With such strong fields, one needs to take into account the vacuum non-linearity as a result of the excitation of virtual electron-positron pairs. 
When the wave propagates in a magnetized plasma, such as the surface of a magnetar, it is expected that  
the wave polarization will be changed. In addition, it has been demonstrated that non-linear effects in strongly magnetized plasmas can generate shock waves, which may contribute to understanding processes in the magnetosphere of neutron stars and the evolution of supernova remnants. Another interesting effect is the photon splitting in magnetars, where one photon decays into two photons due to the presence of non-linear electrodynamics and plasma. Actually, in this case, there is a competition between these two contributions. For more details, we suggest the reviews \cite{Marklund_Shukla_RMP,Ruffini,Uzdensky_Rightley,Gonoskov}.


A lot of effort has been devoted to studying the properties of general non-linear electrodynamics. For example, in the works  \cite{Birula_Birula,Boillat}, the authors obtained the dispersion relations of photon propagation  in an external electromagnetic field. Moreover, the effects of non-linear electrodynamics on the optical properties of vacuum
were also evaluated in refs. \cite{Battesti_Rizzo,Fouche,Zavattini_Valle}.  In this case, it is well-known that the vacuum can be treated as a non-linear optical medium and, consequently, may exhibit optical phenomena such as birefringence, dichroism, photon splitting and wave mixing.  However, in the context of plasma physics, the investigations are mainly restricted to Euler and Kockel electrodynamics. From our understanding, the analysis of plasma waves in a general non-linear electrodynamics has not been carried out in the previous literature. Adopting a unified framework, is useful to analyse how plasma waves will be modified due to non-linear electrodynamics and provide phenomenological results and conditions on the parameters, which enables us to distinguish some non-linear models.


Based on these motivations, we pursue some investigations of modified plasma waves in the context of non-linear electrodynamics. 
Initially, we focus on a general electrodynamics with parity invariance, which encompasses most examples in the literature.  This approach offers the advantage to understand some changes in plasma waves without specifying a particular electrodynamics. Furthermore, in a first attempt at this direction, we consider a cold plasma model, consisting of electrons in a homogeneous ionic background  without the collisional, relativistic and large amplitude effects. At this stage, our main goal is to actually inspect how non-linear electrodynamics interfere in the plasma waves, 
as well as establishing comparisons with the standard modes related to Maxwell electrodynamics. Although a simplified plasma model is examined, one may gain further insights to generalize the results with more involved assumptions.  Indeed, it is expected that the non-linear effects demand strong electromagnetic fields to be detected. Therefore, we have to include additional contributions to apply in the aforementioned astrophysical scenarios or terrestrial laboratories. For instance, in the particular case of circularly polarized waves, we shall consider the ions response, together with the relativistic and large amplitude effects. In this situation, we anticipate that the results may be relevant to investigations with high-intensity laser in plasma, where the approximation of cold plasma is suitable because the thermal motion can be disregarded in comparison with the relativistic motions of electrons and ions. The scope of our work is not to furnish experimental proposals regarding non-linear electrodynamics in plasma. Nevertheless, we believe that our contribution provides some phenomenological results on the refractive index and dispersion relation in a unified framework, which allows us to analyse a set of non-linear models and point out different behaviours.


This paper is organized with the following outline. In Sec. \ref{sec_NL_ED}, we introduce ge\-neral results of non-linear electrodynamics and describe the fluid theory under consideration. Subsequently,  by adopting small amplitude waves and the non-relativistic approximation, we investigate the modified Trivelpiece-Gould dispersion relation, the generalized Appleton-Hartree equation, and their corresponding  principal modes. In Sec. \ref{sec_applic}, we apply the previous results to Euler-Kockel electrodynamics (low-energy quantum effects) and Born-Infeld-type model, as well as present some comparisons with the literature.  After that, in Sec. \ref{sec_L_eff}, the relativistic and large amplitude effects are taken into account for circularly polarized waves in an effective non-linear electrodynamics. We obtain the dispersion relation and analyse some properties of the refractive index. Finally, in Sec. \ref{sec_concl}, our conclusions and perspectives are exhibited. Throughout this work, we adopt the SI units, where $\varepsilon_0$ and $\mu_0$ correspond to the vacuum permittivity and permeability, respectively. In addition, $c= 1/\sqrt{\varepsilon_0 \, \mu_0}$ denotes the speed of light.

\section{Non-linear electrodynamics in cold plasma} \label{sec_NL_ED}

In this section, we present general results of non-linear electrodynamics and the description of the fluid theory.
As already mentioned, the main goal is to investigate some modified plasma waves. The first step in this direction is to discuss the essential features of non-linear electrodynamics. For our purpose, the non-covariant formulation is sufficient. Thus, we begin with the general Lagrangian density 
\begin{equation} \label{def_L}
\mathcal{L} = \frac{1}{\mu_0} \, L_{nl} - \rho \, \phi + {\bf j} \cdot {\bf A} \, ,
\end{equation}
where $\rho$ and ${\bf j}$ denote the charge and current densities, respectively, while $\phi$ and ${\bf A}$ correspond to the electromagnetic potentials such that ${\bf B} = \nabla \times {\bf A}$ and ${\bf E} = - \nabla \phi - \partial {\bf A} / \partial t$.  In order to preserve the Lorentz and gauge symmetries, we have that $L_{nl} = L_{nl}(\mathcal{F}, \mathcal{G})$ must be a function of the invariants 
\begin{equation} \label{def_F_G}
\mathcal{F} = \frac{1}{2} \left( \frac{ {\bf E}^2 }{c^2} - {\bf B}^2 \right) \, \; , \, \; 
\mathcal{G} = \frac{ {\bf E} \cdot {\bf B} }{c} \, .
\end{equation}

The field equations associated with Eq. \eqref{def_L} are given by
\begin{eqnarray}
\nabla \cdot {\bf D} &=& \rho \label{eq_D} \, , \\
\nabla \times {\bf H} &=& {\bf j} + \frac{\partial {\bf D}}{\partial t} \label{eq_H} \, ,
\end{eqnarray}
where we defined the auxiliary fields
\begin{eqnarray}
{\bf D} & = & \frac{\partial L_{nl} }{\partial \mathcal{F} } \, \varepsilon_0 \, {\bf E} + 
\frac{\partial L_{nl} }{\partial \mathcal{G} } \, c \, \varepsilon_0 \, {\bf B} \label{def_D} \, , \\
{\bf H} & = & \frac{\partial L_{nl} }{\partial \mathcal{F} } \, \frac{ {\bf B} }{\mu_0} - 
\frac{\partial L_{nl} }{\partial \mathcal{G} } \, \frac{ {\bf E} }{\mu_0 \, c} \label{def_H} \, .
\end{eqnarray}
For example, in the case of Maxwell electrodynamics, we have $ L_{nl} = \mathcal{F} $, which implies the usual expressions ${\bf D} = \varepsilon_0 \, {\bf E} $ and ${\bf H} =  {\bf B} / \mu_0 $  without the magnetization and polarization vectors.

We highlight that the homogeneous equations remain unchanged, namely,
\begin{equation}
\nabla \times {\bf E} = - \frac{\partial {\bf B}}{\partial t} \, \; , \, \;
\nabla \cdot{\bf B} = 0 \, . \label{Faraday_Gauss}  
\end{equation}

Having made these observations, let us now consider the electromagnetic fields around a uniform and constant magnetic background  field ${\bf B}_0$, such that $ {\bf E} = {\bf \delta E} $ and $ {\bf B} = {\bf B}_0 + {\bf \delta B} $, with ${\bf \delta E}$ and ${\bf \delta B}$ being small perturbations. Keeping this in mind, one can expand Eqs. \eqref{def_D} and \eqref{def_H}, which lead to 
\begin{eqnarray}
 {\bf D} &  \approx  & c_1 \, \varepsilon_0 \, {\bf \delta E} + c_2 \, \varepsilon_0 \, c \left( {\bf B}_0 + {\bf \delta B} \right)
+ d_2 \, \varepsilon_0 \, {\bf B}_0 \left( {\bf B}_0 \cdot {\bf \delta E} \right)
- d_3 \, \varepsilon_0 \, c \, {\bf B}_0 \left( {\bf B}_0 \cdot {\bf \delta B} \right)
\label{D_lin} \, , \\
 {\bf H} & \approx & \frac{c_1}{\mu_0} \, \left( {\bf B}_0 + {\bf \delta B} \right) 
- \frac{c_2}{\mu_0 c} \, {\bf \delta E}  
- \frac{d_1}{\mu_0} \, {\bf B}_0 \left( {\bf B}_0 \cdot {\bf \delta B} \right) 
+ \frac{d_3}{\mu_0 c} \, {\bf B}_0 \left( {\bf B}_0 \cdot {\bf \delta E} \right)
\label{H_lin} \, ,
\end{eqnarray}
where the coefficients $c_1 , c_2 , d_1 , d_2$ and $d_3$ are evaluated at the magnetic background field as follows 
\begin{eqnarray} 
c_{1} &=& \left. \frac{\partial L_{nl}}{\partial{\cal F}}\right|_{{\bf B}_0}
\; \, , \; \,
c_{2} = \left. \frac{\partial L_{nl}}{\partial{\cal G}}\right|_{{\bf B}_0}
\; \, , \; \, \label{coeff_c} \\
d_{1} &=& \left. \frac{\partial^2 L_{nl}}{\partial{\cal F}^2}\right|_{{\bf B}_0}
\; \, , \; \,  
d_{2} = \left. \frac{\partial^2 L_{nl}}{\partial{\cal G}^2}\right|_{{\bf B}_0}
\; \, , \; \,
d_{3} = \left. \frac{\partial^2 L_{nl}}{\partial{\cal F}\partial{\cal G}}\right|_{{\bf B}_0}
\; \, . \label{coeff_d}
\end{eqnarray}
We have adopted a similar notation of ref. \cite{Neves_leo_PRD}. Moreover, from Eq. \eqref{Faraday_Gauss}, we obtain the homogeneous equations for the perturbation fields,
\begin{eqnarray}
\nabla \times {\bf \delta E} &=& - \frac{\partial \, {\bf \delta B} }{\partial t}  \, , \label{Faraday_e_b} \\ 
\nabla \cdot{\bf \delta B} &=& 0 \label{Gauss_b}  \, .
\end{eqnarray}

At this stage, it is important to discuss some subtleties. First of all, using the previous Eqs. \eqref{Faraday_e_b} and \eqref{Gauss_b}, one can show that the coefficient $c_2$ does not contribute to the linearized field equations \eqref{eq_D} and \eqref{eq_H}. Furthermore, we consider non-linear electrodynamics with  parity symmetry (invariance under the discrete transformation ${\bf x} \rightarrow - {\bf x}$) such that
 $L_{nl} = L_{nl}(\mathcal{F}, \mathcal{G})$ need to be invariant under the exchange $\mathcal{G} \rightarrow - \mathcal{G}$. For this reason, we restrict to the models in which  $d_3 = 0$. Thereby, only the coefficients $c_1 , d_1 $ and $ d_2$ will be relevant in our analyses. 


Here it is also opportune to highlight another point of view, namely, one could adopt an effective expansion 
of non-linear electrodynamics in terms of the invariants $\mathcal{F}$ and $\mathcal{G}$, such as
\begin{equation} \label{L_approx}
L_{nl} \approx \sum_{i,j} a_{ij} \, \mathcal{F}^i \mathcal{G}^j \, ,
\end{equation}
with $a_{ij}$ being constants (see, for instance, refs. \cite{Battesti_Rizzo,Fouche}). As expected, the leading-order term  corresponds to the Maxwell contribution $(\mathcal{F})$. From this perspective, we have that $c_1 \approx 1 + \epsilon$,  with $\epsilon << 1$ being a dimensionless parameter and, consequently, $c_1>0$. We shall return to this approach in 
Sec. \ref{sec_L_eff}, where only the quadratic corrections in $\mathcal{F}$ and $\mathcal{G}$ will be investigated.


Next, we pass to describe the fluid theory. We shall adopt a similar methodology of a previous work in ref. \cite{Hass_leo_POP}. 
Initially, we consider a cold plasma in the non-relativistic limit described by 
\begin{eqnarray}
\frac{\partial n}{\partial t} + \nabla \cdot \left( n{\bf u} \right) &=& 0 \, , \label{eq_n} \\
\frac{\partial{\bf u}}{\partial t} + \left( {\bf u} \cdot \nabla \right) {\bf u} &=& 
- \frac{e}{m}({\bf E} + {\bf u}\times{\bf B}) \, , \label{eq_F_L} 
\end{eqnarray}
where $n$ denotes the electrons number density and ${\bf u}$ is the electrons fluid velocity field. In addition, $m$ and $-e$ correspond to the electron mass and charge, respectively.

It should be emphasized that we assume the first order approximation in which ${\bf u} = {\bf \delta u}$ (zero equilibrium fluid velocity), $n = n_0 + \delta n$ and $ {\bf B} = {\bf B}_0 + {\bf \delta B} $, with $n_0$ being the ions background number density, and ${\bf B}_0$ represents the equilibrium magnetic field. Both $n_0$ and ${\bf B}_0$ are considered to be uniform and constant. For the sake of simplicity, we suppose that the ions are infinitely massive, which is suitable for high frequency waves. Furthermore, we shall also disregard the thermal and collisional effects. The inclusion of ions motion and relativistic effects will be carried out in Sec. \ref{sec_L_eff}.

We are now in position to obtain the modified Trivelpiece-Gould dispersion relation and the generalized Appleton-Hartree equation, which will be discussed in the next two subsections.

\subsection{Electrostatic waves} \label{sub_sec_electrostatic}

In this subsection, we investigate the electrostatic waves $({\bf \delta B} = {\bf 0})$ and the corresponding modified Trivelpiece-Gould modes. To accomplish this purpose, let us assume plane wave perturbations  proportional to $ \exp[i({\bf k}\cdot{\bf r} - \omega t)] $, where ${\bf k}$ and $\omega$ are the wave vector and angular wave frequency, respectively. 
From now on, we call the attention that $\delta n , {\bf \delta E}$ and $ {\bf \delta u}$ will denote the Fourier amplitudes.

By considering Eq. \eqref{eq_D} with the charge density $ \rho = e \left( n_0 - n \right) $, as well as the fluid equations \eqref{eq_n} and \eqref{eq_F_L}, we obtain the system
\begin{eqnarray} 
\omega \, \delta n &=& n_0 \, {\bf k} \cdot {\bf \delta u} \, , 
\label{system_1} \\
-i \omega \, {\bf \delta u} &=& - \frac{e}{m} \, \left( {\bf \delta E} + {\bf \delta u} \times {\bf B}_0  \right) \, ,
\label{system_2} \\
i c_1 \, {\bf k} \cdot {\bf \delta E} &+& i d_2 \, {\bf B}_0 \cdot {\bf k} \left( {\bf B}_0 \cdot {\bf \delta E} \right)
= - \frac{e}{\varepsilon_0} \, \delta n \, .
\label{system_3}
\end{eqnarray}

Before proceeding our analysis, we point out that this result also holds for non-linear electrodynamics with $d_3 \neq 0$. Indeed, one can easily check that using ${\bf \delta B} = {\bf 0}$ in Eq. \eqref{D_lin}, the coefficient $d_3$ will not contribute to these expressions. At this stage, the auxiliary field ${\bf H}$ in Eq. \eqref{H_lin} is not required. The magnetic 
field perturbation $({\bf \delta B} \neq {\bf 0})$ will be considered in the next subsection.

For the unmagnetized case $({\bf B}_0 = {\bf 0})$, one can promptly obtain the solution $\omega^2 = \widetilde{\omega}_p^2$, where we adopted the shorthand notation for the modified plasma frequency
\begin{equation} \label{w_tilde_p}
\widetilde{\omega}_p = \omega_p / \sqrt{c_1} \, ,
\end{equation} 
with $ \omega_p = [n_0 \, e^2  /(m \varepsilon_0)]^{1/2} $  being the usual plasma frequency. Here, we remember that $c_1 > 0$, thus $\widetilde{\omega}_p $ is well-defined.

Now, we pass to consider the magnetized case. Let us assume that $ {\bf B}_0 = B_0 \, \hat{z} $ and $ {\bf k} \parallel {\bf \delta E} $ with $ {\bf k} = k \sin\theta \, \hat{x} + k \cos\theta \, \hat{z} $. Solving the system of Eqs. \eqref{system_1}$-$\eqref{system_3}, we arrive at the modified Trivelpiece-Gould dispersion relation
\begin{equation} \label{eq_w4}
\omega^4 - \left( \bar{\omega}_p^2 + \omega_c^2 \right) \, \omega^2 + \bar{\omega}_p^2 \, 
\omega_c^2 \, \cos^{2}\theta = 0 \, ,
\end{equation}
where $ \omega_c = e B_0/m $ corresponds to the electron cyclotron frequency and
\begin{equation} \label{w_bar_p}
\bar{\omega}_p = \frac{1}{\left(  c_1 + d_2 \, B_0^2 \, \cos^2 \theta \, \right)^{1/2}} \, \omega_p \, \; .
\end{equation}

Observe that, for $\theta = \pi/2$ or $d_2=0$, we recover the modified plasma frequency \eqref{w_tilde_p}. 
According to the definition \eqref{coeff_d}, the second condition ($d_2=0$) always occurs for non-linear models that depend only on $L_{nl} = L_{nl}(\mathcal{F})$. 
Furthermore, for a real frequency in Eq. \eqref{w_bar_p}, we have the following constraint 
\begin{equation} \label{constraint}
c_1 + d_2 \, B_0^2 \, \cos^{2} \theta > 0 \, .
\end{equation}

As a consequence, from Eq. \eqref{eq_w4}, we get the solution
\begin{equation} \label{eq_TG}
\omega^2 = \frac{1}{2} \left[ \bar{\omega}_p^2 + \omega_c^2 \pm \left( \left( \bar{\omega}_p^2 - \omega_c^2 \right)^2 + 4 \, \bar{\omega}_p^2 \, \omega_c^2 \, \sin^{2}\theta \right)^{1/2} \right] \, ,
\end{equation}
and, with the condition \eqref{constraint},  one can show that the correspondent modes are always stable $(\omega^2 \geq 0)$. 
Therefore, the standard analysis of electrostatic waves holds with the modified plasma frequency.

Finally, it should be mentioned that, for Maxwell electrodynamics, we have $c_1 = 1$ and $d_1 = d_2 = 0$, which implies $\bar{\omega}_p \rightarrow \omega_p$ and the usual Trivelpiece-Gould dispersion relation is recovered \cite{Chen}.

\subsection{Generalized Appleton-Hartree equation} \label{sub_sec_AH_equation}

In this subsection, we now turn our attention to including the effects of magnetic field perturbation $({\bf \delta B} \neq {\bf 0})$.  Here, we will analyse non-linear electrodynamics with parity invariance such that $d_3 = 0$. 
As done before, we consider plane wave perturbations proportional to $ \exp[i({\bf k}\cdot{\bf r} - \omega t)] $.
Additionally, we also assume that $ {\bf B}_0 = B_0 \, \hat{z} $ and $ {\bf k} = k \sin\theta \, \hat{x} + k \cos\theta \, \hat{z} $.


Following the usual procedure \cite{Swanson,Stix}, we first manipulate Eq. \eqref{eq_F_L}
to isolate the linearized velocity in terms of electric amplitude and frequency, which yields
\begin{equation} \label{velocities} 
\delta u_x = \frac{e}{m} \, \frac{\left( \omega_c \, \delta E_y + i \omega \, \delta E_x \right) }{\left( \omega_c^2 
- \omega^2 \right)} 
\, , \,
\delta u_y = \frac{e}{m} \, \frac{\left( - \omega_c \, \delta E_x + i \omega \, \delta E_y \right)}{ \left( \omega_c^2 
- \omega^2 \right)}
\, , \,
\delta u_z = - \frac{i  e}{m \omega} \, \delta E_z 
\, . 
\end{equation}
Subsequently, from Eq. \eqref{Faraday_e_b}, one can easily see that $ {\bf \delta B} =  {\bf k} \times {\bf \delta E} / \omega $. By inserting these results into the modified Amp\`ere-Maxwell law \eqref{eq_H} with the current density ${\bf j} = - n e \, {\bf u}$, we finally arrive at the system
\begin{equation} \label{matrix_E}
\begin{bmatrix}
   {\cal S} - \eta^2  \cos^{2}\theta &
   -i {\cal D} & \eta^2 \, \cos\theta  \sin\theta \\
   i {\cal D} & {\cal S} - \eta^2 \,  \chi(\theta) & 0 \\
	\eta^2  \cos\theta  \sin\theta & 0 & {\cal P} - \eta^2 \sin^{2}\theta 
\end{bmatrix}
\begin{bmatrix} 
	\delta E_x \\ \delta E_y \\ \delta E_z 
\end{bmatrix} = 0 \, , 
\end{equation}
where we defined the modified Difference $({\cal D})$, Sum $({\cal S})$  and Plasma $({\cal P})$  coefficients,
\begin{equation} \label{plasma_coeff}
{\cal D} = \frac{\omega_c \, \widetilde{\omega}_p^2}{\omega(\omega_c^2-\omega^2)} \, , \, \quad {\cal S} = 1 + \frac{\widetilde{\omega}_p^2}{\omega_c^2-\omega^2} \, , \, \quad {\cal P} =  1 - \frac{\widetilde{\omega}_p^2}{\omega^2} + \frac{d_2 }{c_1} \, B_0^2 \, .
\end{equation}
with $ \widetilde{\omega}_p = \omega_p / \sqrt{c_1} $ being the modified plasma frequency. In addition, 
$ \eta = ck/\omega $ denotes the refractive index and 
\begin{equation} \label{chi_theta}
 \chi(\theta) = 1 - \frac{d_1}{c_1} \, B_0^2 \sin^2 \theta \, .
\end{equation}

The previous definitions clearly show the contributions of non-linear electrodynamics. First of all, we adopted the standard definitions for the coefficients ${\cal D}$ and ${\cal S}$ with $\omega_p \rightarrow\widetilde{\omega}_p $. However, for the   coefficient ${\cal P}$, we have a non-trivial contribution of $d_2 B_0^2/c_1$. Furthermore, we would like to point out the angular dependence in Eq. \eqref{chi_theta} due to the presence of $d_1 B_0^2/c_1$.

The non-trivial solutions of Eq. \eqref{matrix_E} are obtained by imposing that the determinant of the matrix must vanish, which leads to
\begin{equation} \label{eta_eq}
A \, \eta^4 - B \, \eta^2 + C = 0 \, , 
\end{equation}
where the coefficients are read below
\begin{eqnarray} 
A &=&  \chi(\theta) \, \left[ {\cal S} \sin^{2}\theta +  {\cal P}  \cos^{2} \theta \right] \, , 
\label{def_A} \\
B &=& {\cal R} {\cal L} \, \sin^{2}\theta +  {\cal S} {\cal P}  \, [ \chi(\theta) +  \cos^{2}\theta] \, , 
\label{def_B} \\
C &=&  {\cal P R L} \, ,
\label{def_C} 
\end{eqnarray}
in which we have also introduced $ {\cal R} = {\cal S} + {\cal D} $ and $ {\cal L} = {\cal S} - {\cal D} $ for the modified Right and Left coefficients, respectively.

After some standard manipulations of Eq. \eqref{eta_eq}, we promptly find that
\begin{equation}
\eta^2 = 1 - \,\frac{2(A-B+C)}{2A-B \pm \sqrt{B^2-4 A C}} \, \; ,
\end{equation}
and, by substituting the coefficients \eqref{def_A}$-$\eqref{def_C}, this expression can be written as
\begin{equation} \label{A_H_eq_gen}
\eta^2 = 1 - \frac{\widetilde{\omega}_p^2/\omega^2}{Q} \,, \quad  Q = \left( Q_0 \pm F \right)/Q_1 
\, , \end{equation} 
with $F^2 \equiv B^2 - 4 AC$ given by
\begin{eqnarray} \label{def_F}
F^2 &=& \left[ {\cal R} {\cal L} - \left( 1 + \frac{d_1}{c_1} \, B_0^2 \right) {\cal S} {\cal P} \right]^2 \sin^4 \theta +
4 \left[  1 -  \chi(\theta) \right] {\cal S}^2 {\cal P}^2 +
4 \left[  1 -  \chi(\theta) \right] {\cal S} {\cal P} {\cal R} {\cal L} \sin^2 \theta \nonumber \\
&-& 4 \left[  1 -  \chi(\theta) + \frac{d_1}{c_1} \, B_0^2 \right] {\cal S}^2 {\cal P}^2 \sin^2 \theta +
4 \,  \chi(\theta) \, {\cal P}^2 {\cal D}^2 \cos^2 \theta
\, , \end{eqnarray} 
and the definitions
\begin{eqnarray} \label{def_Q0}
Q_0 &=& \left( 1 - \frac{\widetilde{\omega}_p^2}{\omega^2} + \frac{d_2}{c_1} B_0^2 \right)
\left[  \chi(\theta) - 1 - ( \chi(\theta) +1) \frac{\widetilde{\omega}_p^2}{\omega_c^2 - \omega^2} \right] + \nonumber \\
&+& \frac{\widetilde{\omega}_p^2}{\omega^2} \, \sin^2 \theta \left[ 
(2  \chi(\theta)-1) \frac{\omega_c^2}{\omega_c^2 - \omega^2} 
- (2 \chi(\theta) -1) \frac{d_2}{c_1} \, B_0^2 \, \frac{\omega^2}{\widetilde{\omega}_p^2} 
+  \frac{d_2}{c_1} \, B_0^2 \, \frac{\omega^2}{\omega_c^2 - \omega^2 }
\right] \, , \end{eqnarray} 
\begin{eqnarray} \label{def_Q1}
Q_1 &=&  \frac{ 2\left(  \omega^2 - \widetilde{\omega}_p^2 + d_2 \, B_0^2 \, \omega^2/c_1 \right) }{\omega_c^2 - \omega^2} 
\left[ 1 - \chi(\theta) - \frac{\widetilde{\omega}_p^2}{\omega^2} \right] + \nonumber \\
&+& 2 \sin^2 \theta \left[  ( \chi(\theta) - 1) \frac{\omega_c^2}{\omega_c^2 - \omega^2} 
- ( \chi(\theta) - 1) \, \frac{\omega^2}{\widetilde{\omega}_p^2} \, \frac{d_2}{c_1} \, B_0^2 +
\frac{\omega^2}{\omega_c^2 - \omega^2} \, \frac{d_2}{c_1} \, B_0^2
\right] \, . \end{eqnarray} 

We highlight that Eq. \eqref{A_H_eq_gen} corresponds to the generalized Appleton-Hartree equation with the contributions of non-linear electrodynamics. As expected, in the Maxwell case ($c_1 = 1$ and $d_1 = d_2 = 0$), we recover the well-known Appleton-Hartree equation \cite{Swanson,Stix}, where 
\begin{equation}
Q = 1 - \frac{\omega_c^2 \, \sin^{2}\theta}{2 (\omega^2 - \omega_p^2)} \pm \left(\frac{\omega_c^4 \, \sin^{4}\theta}{4 (\omega^2 -\omega_p^2)^2} + \frac{\omega_c^2 \, \cos^{2}\theta}{\omega^2}\right)^{1/2} \, . 
\end{equation}


The general solution of Eq. \eqref{A_H_eq_gen} is quite involved. However, the analyses of the principal modes are feasible. In what follows, we consider the situations of propagation parallel or perpendicular to the equilibrium magnetic field ${\bf B}_0$.

For $\theta = 0$, we obtain $ \chi(\theta) =1$ and $Q = 1 \pm \omega_c/\omega$, such that Eq. \eqref{A_H_eq_gen} can be written in the two modes below
\begin{eqnarray}
\frac{c^2 k^2}{\omega^2} &=& 1 - \frac{\widetilde{\omega}_p^2}{\omega ( \omega - \omega_c)} \label{RCP} \, , \\
\frac{c^2 k^2}{\omega^2} &=& 1 - \frac{\widetilde{\omega}_p^2}{\omega ( \omega + \omega_c)} \label{LCP} \, ,
\end{eqnarray}
which correspond to the modified right-hand and left-hand circularly polarized waves (RCP and LCP), respectively.
Therefore, only the coefficient $c_1$ will contribute to changes in the RCP and LCP modes and the well-known properties are reproduced here through the replacement of plasma frequency $\omega_p \, \rightarrow \,  \widetilde{\omega}_p = \omega_p / \sqrt{c_1}$ in the usual results \cite{Chen,Swanson,Stix}.

Moreover, the parallel propagation case allows another solution. By using $\theta = 0$, all the coefficients $A,B$ and $C$ in Eqs. \eqref{def_A}$-$\eqref{def_C} become proportional to ${\cal P}$. Hence, ${\cal P} = 0$ is a possible solution of Eq. \eqref{eta_eq}, which leads to 
\begin{equation} \label{P_sol}
\omega = \frac{\omega_p}{ \sqrt{  c_1 + d_2 \, B_0^2 } }  \; .
\end{equation}
Note that this frequency coincides with $\bar{w}_p$ defined in the context of electrostatic waves (see Eq. \eqref{w_bar_p} with $\theta = 0$). 

\begin{figure*}[th]
\centering
\includegraphics[width=0.5\textwidth]{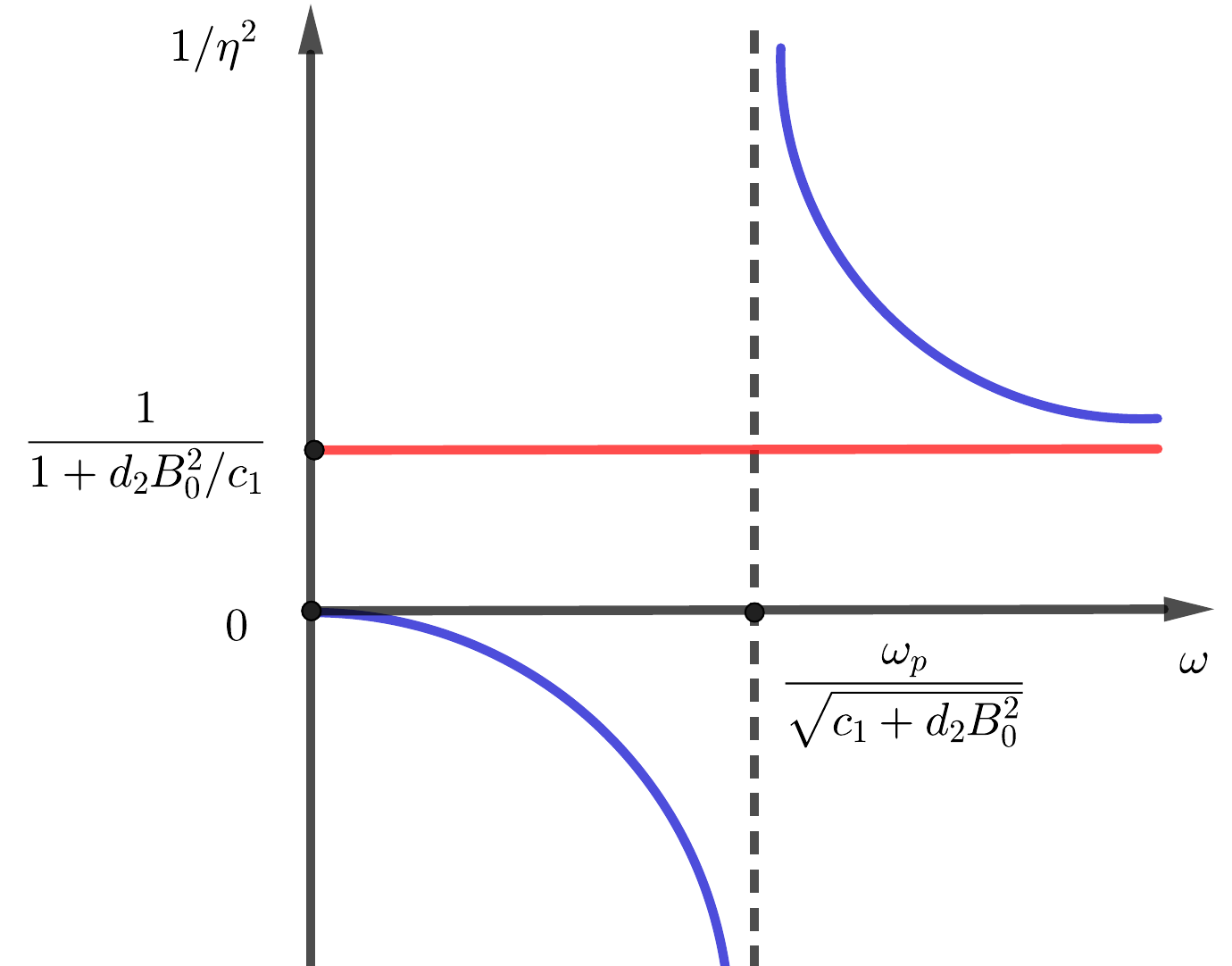}
\caption{Dispersion relation of the modified $O-$mode from Eq. \eqref{O_mode}. A cut-off frequency occurs at $\omega_p/\sqrt{c_1 + d_2 B_0^2}$ and the wave does not propagate in the region $0 < \omega < \omega_p/\sqrt{c_1 + d_2 B_0^2}$. }
\label{fig-modo-O}
\end{figure*}

Next, we pass to consider Eq. \eqref{A_H_eq_gen} with $\theta = \pi/2$. After algebraic manipulations, we obtain two solutions. The first one describes the modified ordinary $(O)$ mode, given by 
\begin{equation} \label{O_mode}
\omega^2 = \frac{c^2 k^2 + \widetilde{\omega}_p^2}{\left( 1 + \frac{d_2}{c_1} B_0^2 \right)} \, .
\end{equation}
In comparison with the usual $O-$mode $(\omega^2 = c^2 k^2 + \omega_p^2)$, we get new contributions due to the coefficients $c_1$ and $d_2$. The analysis of the modified $O-$mode can be carried out with the definitions of cut-off and resonance, which help us to divide the regions of propagation and non-propagation. We remember that a cut-off occurs whenever the refractive index $\eta \rightarrow 0$, while a resonance happens when $\eta \rightarrow \infty$. In general, the wave will be reflected at a cut-off and absorbed at a resonance. A qualitative description is displayed in Fig. \ref{fig-modo-O}, where we consider a diagram $1/\eta^2$ versus $\omega$. In this case, there is one cut-off at $\omega_p/\sqrt{c_1 + d_2 B_0^2}$ and no resonances. The wave only propagates in the region with $\eta^2 >0 $, namely, for frequency $\omega > \omega_p/\sqrt{c_1 + d_2 B_0^2} $. 
At this stage, we recall that $1/\eta^2 = v_\phi^2/c^2$, where $v_\phi$ denotes the phase velocity. Therefore, the wave
travels faster or slower than $c$ depending on the values of the coefficients $c_1$ and $d_2$ (the red line may be above or below to $1/\eta^2 =1$). For high frequency, we have that $1/\eta^2$ approaches to $1/(1+ d_2 B_0^2/c_1)$.

The other solution for perpendicular propagation corresponds to the modified extraordinary $(X)$ mode, described by
\begin{equation} \label{X_mode}
\frac{c^2 k^2}{\omega^2} \left( 1 - \frac{d_1}{c_1}  B_0^2 \right) = 
1 - \frac{\widetilde{\omega}_p^2}{\omega^2} \, 
\frac{(\omega^2 - \widetilde{\omega}_p^2) }{(\omega^2 - \widetilde{\omega}_h^2)} \, \; ,
\end{equation}
with $\widetilde{\omega}_h$ being the modified upper-hybrid frequency such that
$ \widetilde{\omega}_h^2 = \omega_c^2 + \widetilde{\omega}_p^2 $. By comparing with the usual $X-$mode, 
\begin{equation} \label{X_usual_mode}
\frac{c^2 k^2}{\omega^2}  = 
1 - \frac{\omega_p^2}{\omega^2} \, 
\frac{(\omega^2 - \omega_p^2) }{(\omega^2 - \omega_h^2)} \, \; ,
\end{equation}
we observe that only the coefficients $c_1$ and $d_1$ will introduce modifications by means of a factor $(1 - d_1 B_0^2/c_1)$ and the replacements $\omega_p \rightarrow \widetilde{\omega}_p$ and $\omega_h \rightarrow \widetilde{\omega}_h$.

It is worthwhile mentioning that Eq. \eqref{X_mode} can be recast as
\begin{equation} \label{X_mode_1}
\frac{c^2 k^2}{\omega^2} \left( 1 - \frac{d_1}{c_1}  B_0^2 \right) = 
\frac{(\omega^2 - \widetilde{\omega}_L^2) (\omega^2 - \widetilde{\omega}_R^2) 
}{ \omega^2 (\omega^2 - \widetilde{\omega}_h^2)} \, \; ,
\end{equation}
where we define the modified cut-off frequencies
\begin{eqnarray} 
\widetilde{\omega}_L &=& \frac{1}{2} \left[  - \omega_c + \sqrt{\omega_c^2 + 4 \, \widetilde{\omega}_p^2 }  
\, \right] \label{wl} \, , \\ 
\widetilde{\omega}_R &=& \frac{1}{2} \left[  \omega_c + \sqrt{\omega_c^2 + 4 \, \widetilde{\omega}_p^2 }  
\, \right] \label{wr} \, .
\end{eqnarray}

\begin{figure*}[th]
\centering
\includegraphics[width=0.5\textwidth]{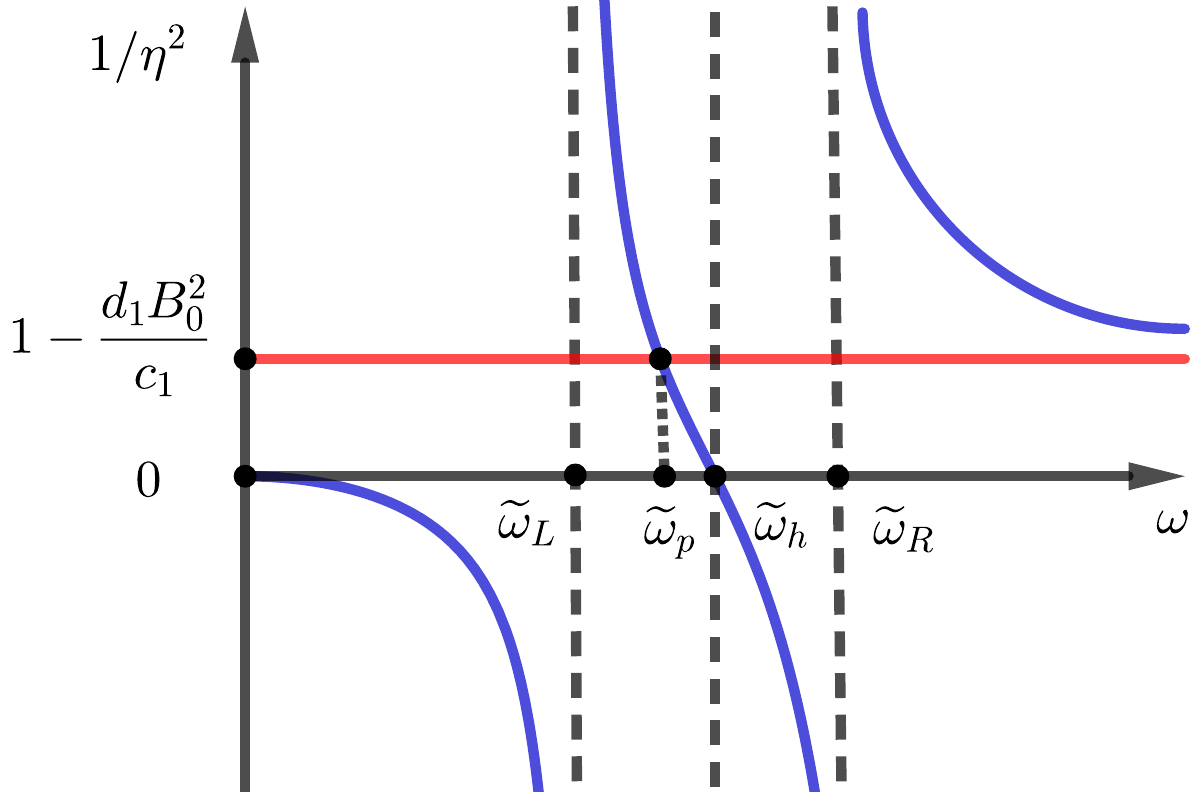}
\caption{Dispersion relation of the modified $X-$mode from Eq. \eqref{X_mode_1}. A resonance happens at $\widetilde{\omega}_h$ and the cut-off frequencies are located at $\widetilde{\omega}_L$ and $\widetilde{\omega}_R$. Forbidden bands are settled in $0 < \omega < \widetilde{\omega}_L$ and $\widetilde{\omega}_h < \omega < \widetilde{\omega}_R$. }
\label{fig-modo-X}
\end{figure*}

From these results, one can easily understand the behaviour of the modified $X-$mode, which is exhibited in Fig. \ref{fig-modo-X}. First of all, there is a resonance at $\widetilde{\omega}_h$, and the cut-off frequencies are situated at $\widetilde{\omega}_L$ and $\widetilde{\omega}_R$. In the regions $0 < \omega < \widetilde{\omega}_L$ and $\widetilde{\omega}_h < \omega <\widetilde{\omega}_R$, we see that $1/\eta^2$ is negative and, consequently, there is no propagation. On the other hand, we have two regions of propagation, given by $ \widetilde{\omega}_L < \omega < \widetilde{\omega}_h$ and $ \omega > \widetilde{\omega}_R $, which are separated by a stop band. Furthermore, it should be mentioned that the coefficients $c_1$ and $d_1$ also modify the regions in which the wave travels faster or slower than $c$ (again, the red line may be above or below to $1/\eta^2 =1$). For high frequency, we have that $1/\eta^2$ approaches to $(1 - d_1 B_0^2/c_1)$.


A final comment we would like to raise concerns the limit of a null plasma frequency. From Eqs. \eqref{RCP} and \eqref{LCP}, we find the trivial solution $\omega = ck$. In this limit, it is also expected to recover the well-known dispersion relations of non-linear electrodynamics in vacuum with an external magnetic field. Indeed, from Eqs. 
\eqref{O_mode} and \eqref{X_mode}, one can promptly arrive at
\begin{eqnarray}
\omega & \approx & \frac{c k}{ \sqrt{ 1 + \frac{d_2}{c_1} B_0^2 }} \label{w1_vacuum} \, , \\
\omega & \approx & c k \, \sqrt{ 1 - \frac{d_1}{c_1} B_0^2 } \label{w2_vacuum} \, , 
\end{eqnarray}
which agrees with the results in refs. \cite{Birula_Birula,Neves_leo_PRD}. Notice that the dispersion relations \eqref{w1_vacuum} and \eqref{w2_vacuum} can be obtained from the asymptotic behaviour for high frequency in Figs. \ref{fig-modo-O} and \ref{fig-modo-X}, respectively. 


Having established the principal modes, we are now equipped to investigate some specific models. 
It will be done in the next section.

\section{  Particular non-linear electrodynamics} \label{sec_applic}

So far we have not specified any electrodynamics, except for some considerations about the Maxwell limit. The only assumption is parity invariance so that we restrict to the cases with $d_3 =0$. In what follows, we apply our results to particular models. Initially, we consider the well-known Euler-Kockel electrodynamics to illustrate our methodology and establish some comparisons with the literature. After that, we pursue an investigation of a generalized Born-Infeld electrodynamics, which has been a subject of intense research.

\subsection{Euler-Kockel electrodynamics} \label{sub_sec_EH}

The effective theory obtained by Euler and Kockel (EK) \cite{Euler_Kockel} is one of the most investigated non-linear electrodynamics. This proposal takes into account the low-energy quantum effects of vacuum polarization produced by virtual electron$-$positron pair, which leads to the following corrections
\begin{equation} \label{L_EK}
L_{EK} = \mathcal{F} + \frac{2}{45} \, \frac{\alpha^2}{m^4} \, \frac{\hbar^3 \varepsilon_0}{c^3}
\left( 4  \mathcal{F}^2 + 7 \mathcal{G}^2 \right) \, ,
\end{equation}
where $\hbar \equiv h/2\pi$ denotes the reduced Planck constant and $\alpha \equiv e^2/ (4 \pi \varepsilon_0 \hbar c) \approx 1/137 $ is the fine structure constant.

It is appropriate to remark that the EK model holds for frequencies below the Compton frequency $( \omega << \omega_e = m c^2/\hbar$), and for  electromagnetic fields in the weak regime $ |{\bf E}| << E_c $ and $ |{\bf B}| << B_c $, where the critical fields are given by 
\begin{equation} \label{critical_E_B}
E_c \equiv m^2 c^3/e \hbar \approx 1,3 \times 10^{18} \, \textrm{V/m} \; \, , \; \,
B_c \equiv m^2 c^2/e \hbar \approx 4,4 \times 10^9 \, \textrm{T} \, .
\end{equation} 

Mention should be made that EK result was genera\-lized by Heisenberg and Euler (HE) electrodynamics  \cite{Heisenberg_Euler}, where the authors obtained a non-perturbative expression including high-order  quantum corrections. For more details on EK and HE electrodynamics, we point out the reviews \cite{Dunne_1,Dunne_2} and references therein. Here, let us focus on the leading quantum correction, as described in Eq. \eqref{L_EK}.

According to the prescription in Eqs. \eqref{coeff_c} and \eqref{coeff_d} with $L_{nl} = L_{EK}$, we find that
\begin{eqnarray}
c_1 &=& 1 - \frac{2}{45} \, \frac{\alpha}{\pi} \, \left( \frac{B_0}{B_c} \right)^2 \,  
\, \; , \label{c1_EK} \\
d_1 &=& \frac{4}{45} \, \frac{\alpha}{\pi} \, \frac{1}{B_c^2} 
\, \; , \label{d1_EK} \\
d_2 &=& \frac{7}{45} \, \frac{\alpha}{\pi} \, \frac{1}{B_c^2}
\, \; . \label{d2_EK} 
\end{eqnarray}
As already expected due to parity symmetry, we also obtain $d_3=0$. From Eq. \eqref{c1_EK}, one can immediately see that $c_1>0$.
In this model, note that the coefficients $d_1$ and $d_2$ do not depend on the equilibrium magnetic field $B_0$. Moreover, we draw attention to the fact that $d_2>0$, then the constraint \eqref{constraint} related to electrostatic waves is automatically satisfied for any $\theta$-angle with the modified plasma frequency
\begin{equation} \label{EK_electrostatic}
\bar{\omega}_p \approx \left[ 1 + \frac{\alpha}{45 \pi} \, \left( \frac{B_0}{B_c} \right)^2 \, 
\left( 1 - \frac{7}{2} \cos^2 \theta \right) \right] \omega_p
\, . \end{equation}


Next, let us examine the RCP mode. As described in Eq. \eqref{RCP}, only $c_1$ will be relevant. Therefore, using  the coefficient \eqref{c1_EK} and the assumption $B_0 << B_c$, we obtain
\begin{equation} \label{RCP_EK}
\eta^2 \approx 1 - \frac{\omega_p^2}{\omega (\omega - \omega_c) } 
- \frac{2}{45} \frac{\alpha}{\pi} \left( \frac{B_0}{B_c} \right)^2 
\frac{\omega_p^2}{\omega (\omega - \omega_c) } 
\; \, , \end{equation}
which agrees with the result of ref. \cite{Lundin_RCP_mode}. Similarly, one can get the LCP mode by substituting 
$\omega_c \rightarrow - \omega_c$ in the previous expression.


We now consider the $O-$mode. In this case, it is opportune to define the dimensionless parameter $\xi = (\alpha/90 \pi) (B_0/B_c)^2$, such that the coefficients \eqref{c1_EK} and \eqref{d2_EK} can be written as $ c_1 = 1 - 4 \xi $ and $ d_2 B^2_0 = 14 \xi $. Keep this definition in mind and  $\xi << 1$, from Eq. \eqref{O_mode}, one can promptly arrive at
\begin{equation} \label{O_mode_EK}
\omega^2 \approx \left( 1 - 14 \xi \right) \, c^2 k^2 + \left( 1 - 10 \xi  \right) \, \omega_p^2 
\; \, , \end{equation}
which is consistent with refs. \cite{Brodin_NjP,Brodin_PRL}.


\subsection{Born-Infeld-type electrodynamics} \label{sub_sec_BI}

One of the first non-linear electrodynamics was proposed by Born and Infeld (BI) to remove the singularity of the electric field at short distances and, consequently, to avoid the divergence of the electron self-energy \cite{Born_Infeld}. The BI model is described by
\begin{equation} \label{L_BI}
L_{BI} = \frac{\beta^2}{c^2} \left[ 1 - \sqrt{
1 - \frac{2 c^2}{\beta^2} \, \mathcal{F} - \frac{c^4}{\beta^4} \, \mathcal{G}^2 } \, \right] \, , 
\end{equation}
where $\beta$ denotes the BI parameter with the same dimension as the electric field.

Observe that, at the limit of extremely large $\beta$, this model reduces to Maxwell electrodynamics. Interestingly, the BI model was recovered in the low-energy limit of string theories \cite{Fradkin,Bergshoeff}. Another particular feature is the absence of the birefringence phenomenon under an external electromagnetic field \cite{Birula}. However, it is pertinent to comment that this result can be circumvented by coupling the BI model with a dark matter candidate \cite{Paixao}.

In this subsection, based on ref. \cite{Kruglov_MPLA}, we address to the following generalization
\begin{equation} \label{L_BI_p}
L_{BI-\textrm{type}} =  \frac{\beta^2}{c^2} \left[ 1 - \left(
1 - \frac{ c^2}{p \beta^2} \, \mathcal{F} - \frac{c^4}{2p \beta^4} \,  \zeta \, \mathcal{G}^2
\right)^p \, \right] \, , 
\end{equation}
with $p$ and $ \zeta$ being dimensionless parameters. One can immediately see that the BI model (\ref{L_BI}) is recovered when $p=1/2$ and $ \zeta =1$. Similar models were also analysed in refs. \cite{Neves_leo_PRD,BI_type,Neves_BI_EW,Kruglov_JPA}, where the authors discuss the effects of birefringence, dichroism and obtain a finite electric field at the origin for some specific values of the parameters.


It is important to point out that at the limit of low-energy fields $(\beta >> c \, \mathcal{F}  ,  c \, \mathcal{G})$, the Born-Infeld-type model (\ref{L_BI_p}) can be approximated by
\begin{equation} \label{L_BI_approx}
L_{BI-\textrm{type}} \approx  \mathcal{F} + \frac{(1-p)}{p} \, \frac{c^2}{\beta^2} 
\frac{\mathcal{F}^2}{2} +  \zeta \, \frac{c^2}{\beta^2} \frac{\mathcal{G}^2}{2} \, , 
\end{equation}
which corresponds to the most general post-Maxwellian model up to second order in the invariants $\mathcal{F}$ and $\mathcal{G}$. 
Notice that the term proportional to $\mathcal{F} \mathcal{G}$ does not appear due to parity symmetry.

In addition, by taking the limit $p \rightarrow \infty$ in Eq. (\ref{L_BI_p}), we arrive at the so-called exponential electrodynamics,
\begin{equation} \label{L_exp}
L_{\textrm{exp}} = \lim_{p \rightarrow \infty} L_{BI-\textrm{type}} =
\frac{\beta^2}{c^2} \left[  1 -  \textrm{exp} \left( - \frac{c^2}{\beta^2} \, \mathcal{F} 
 - \frac{c^4}{\beta^4} \, \frac{ \zeta \, \mathcal{G}^2}{2}
\right) \right] \, .
\end{equation}
This type of model was initially investigated in the context of black hole solutions \cite{Hendi_1,Hendi_2}. For a detailed review of BI-type and exponential electrodynamics, as well as other generalizations and their properties, we highlight the recent work of ref. \cite{Dehghani}.

Therefore, the BI-type model (\ref{L_BI_p}) allows us to investigate a series of electrodynamics in the literature. We only need to consider different values and limits in the parameter space  $(p, \zeta ,\beta)$.


After these motivations, we now proceed to study the corresponding cold plasma waves. First of all, using the prescription in Eqs. (\ref{coeff_c}) and (\ref{coeff_d}) with $L_{nl} = L_{BI-\textrm{type}}$, we obtain
\begin{eqnarray}
c_1 &=&  \left(  1 + \frac{c^2 B_0^2}{2p \, \beta^2} \, \right)^{p-1}
\, \; , \label{c1_BI} \\
d_1 &=& \frac{(1-p)}{p} \, \frac{c^2}{\beta^2} \, \left(  1 + \frac{c^2 B_0^2}{2p \, \beta^2} \, \right)^{p-2}
\, \; , \label{d1_BI} \\
d_2 &=&  \zeta \, \frac{c^2}{\beta^2} \, \left(  1 + \frac{c^2 B_0^2}{2p \, \beta^2} \, \right)^{p-1}
\, \; . \label{d2_BI} 
\end{eqnarray}
Recalling again that we get $d_3=0$. 

The requirement that $c_1 > 0$ is satisfied by $p > - c^2 B_0^2/(2 \beta^2)$. However, from Eq. \eqref{L_BI_p}, we  note that $p<0$ would allow field configurations in which $L_{BI-\textrm{type}} $ may diverge. To avoid such singularities, we restrict ourselves to cases where $p>0$.

We begin our analysis with electrostatic waves. Before going into details, it should be mentioned that the  BI model was investigated in ref. \cite{Burton_JPA}, but the authors focused on large amplitude effects and the  maximum values for the electric field and frequency. Here, as already emphasized, we will be interested in the modified Trivelpiece-Gould modes \eqref{eq_TG}. For this purpose, we  need to figure out the modified plasma frequency \eqref{w_bar_p}. In this manner, by using the coefficients \eqref{c1_BI} and \eqref{d2_BI}, we obtain
\begin{equation} \label{bar_w_BI}
\bar{\omega}_p = \omega_p \,  \left( 1 + \frac{c^2 B_0^2}{2p \, \beta^2}  \right)^{(1-p)/2} 
\left( 1 +  \zeta \, \frac{c^2 B_0^2}{\beta^2} \,  \cos^2 \theta \right)^{-1/2}
\, . \end{equation}
As a consequence, we observe that negative values for $ \zeta$ are possible with the condition $| \zeta| \cos^2 \theta < \beta^2/(c^2 B_0^2)$. For $ \zeta \geq 0$, we have $d_2 \geq 0$ and the constraint \eqref{constraint} is contemplated for any $\theta$-angle.

It is also interesting to consider the weak field approximation $\beta^2 >> c^2 B_0^2$. Therefore, we can expand Eq. \eqref{bar_w_BI}, which leads to
\begin{equation} \label{bar_w_BI_approx}
\bar{\omega}_p \approx \omega_p - \omega_p \, \frac{c^2 B_0^2}{2 \beta^2}  
\left[ \frac{(p-1)}{2p} +  \zeta \cos^2 \theta \right]
\, . \end{equation}
This result clearly shows that $\bar{\omega}_p \approx \omega_p$ for some regions in the parameter space  $(p, \zeta,\theta)$. In other words, the first order contributions from non-linear electrodynamics cancel out if the parameters satisfy $(1-p)/2p =  \zeta \cos^2 \theta$. For instance, we display specific values in table \ref{tab}. Notice that the Born-Infeld model $(p=1/2 \, \; \textrm{and} \, \;  \zeta =1)$ obeys this condition only for $\theta = \pi/4$. In the case of exponential electrodynamics $(p \rightarrow \infty)$, the cancellation occurs for particular values of the parameter $ \zeta$. Moreover, for $p=1$ and perpendicular propagation $(\theta = \pi/2)$, the condition is automatically satisfied for any $ \zeta << \beta^2/(c^2 B_0^2)$.


\begin{table}[h]\centering 
  \begin{tabular}{ | c | c | c | c | c |}
    \hline
   \, $\theta$ \, & 0 & \, $\pi/6$ \, & \, $\pi/4$ \, & \, $\pi/3$ \, \\ \hline
   \, $ \zeta$ \, & \, $ \frac{1-p}{2p} $ \, & \, $ \frac{2}{3} \frac{(1-p)}{p} $ \, & \, $ \frac{1-p}{p} $ \, 
   & \, $ 2\frac{(1-p)}{p} $ \,  \\ \hline 
  \end{tabular}
  \caption{Specific values for $(\theta,  \zeta)$ where $\bar{\omega}_p \approx \omega_p$ in the weak field regime.}
  \label{tab} 
\end{table}


Our next undertaking is the circularly polarized waves. Let us consider the RCP mode. In this case, only the coefficient $c_1$ will be pertinent. According to Eqs. \eqref{RCP} and \eqref{c1_BI}, we promptly arrive at
\begin{equation} \label{RCP_BI-type}
\eta^2 = 1 - \frac{\omega_p^2}{\omega (\omega - \omega_c)} \left( 1 + \frac{c^2 B_0^2}{2p \, \beta^2} \right)^{1-p}
\, .\end{equation}
In the same way, one can obtain the LCP mode by replacing $\omega_c \rightarrow - \omega_c$ in the previous expression.
For the particular case of BI electrodynamics with the notation $\kappa = 1/\beta$, Eq. \eqref{RCP_BI-type} reduces to
\begin{equation} \label{RCP_BI}
(\omega - \omega_c) \left( \omega - \frac{c^2 k^2}{\omega} \right) = 
\omega_p^2 \, \sqrt{1 + \kappa^2 c^2 B_0^2}
\, , \end{equation}
and we recover the result in ref. \cite{Burton_SPIE}.

As before, we consider the weak field approximation, such that Eq. \eqref{RCP_BI-type} takes the form
\begin{equation} \label{RCP_BI-type_app}
\eta^2 \approx 1 - \frac{\omega_p^2}{\omega (\omega - \omega_c)} 
- \frac{(1-p)}{2p} \, \frac{c^2 B_0^2}{\beta^2} \, \frac{\omega_p^2}{\omega (\omega - \omega_c)} 
\, .\end{equation}
This expression is very similar to the one obtained for EK electrodynamics in Eq. \eqref{RCP_EK}. However, we have an additional possibility here, namely, the last term in Eq. \eqref{RCP_BI-type_app} can assume positive or negative values depending on whether $p>1$ or $0<p<1$.

We now pass to investigate the modified $O-$mode. By substituting the coefficients \eqref{c1_BI} and \eqref{d2_BI} in Eq. \eqref{O_mode}, we find that

\begin{equation} \label{modeO_BI-type}
\eta^2 = 1 +  \zeta \, \frac{c^2 B_0^2}{\beta^2} 
- \left( 1 + \frac{c^2 B_0^2}{2p \, \beta^2} \right)^{1-p} \, \frac{\omega^2_p}{\omega^2}
\; \, . \end{equation}

At this stage, we remember the discussion related to Fig. \ref{fig-modo-O}. The coefficients $c_1$ and $d_2$ modify the region where the wave may travel faster or slower than $c$, according to the horizontal red line defined by $1/(1+ d_2 B_0^2/c_1)$. In the case of BI-type model, this line is given by $1/(1 +  \zeta \, c^2 B_0^2/\beta^2)$. Therefore, we need to analyse the situations with $ \zeta > 0$ and $ \zeta < 0$, as well as specific values for $p$ and $c^2 B_0^2/\beta^2$.

\begin{figure*}[th]
\centering
\includegraphics[width=0.5\textwidth]{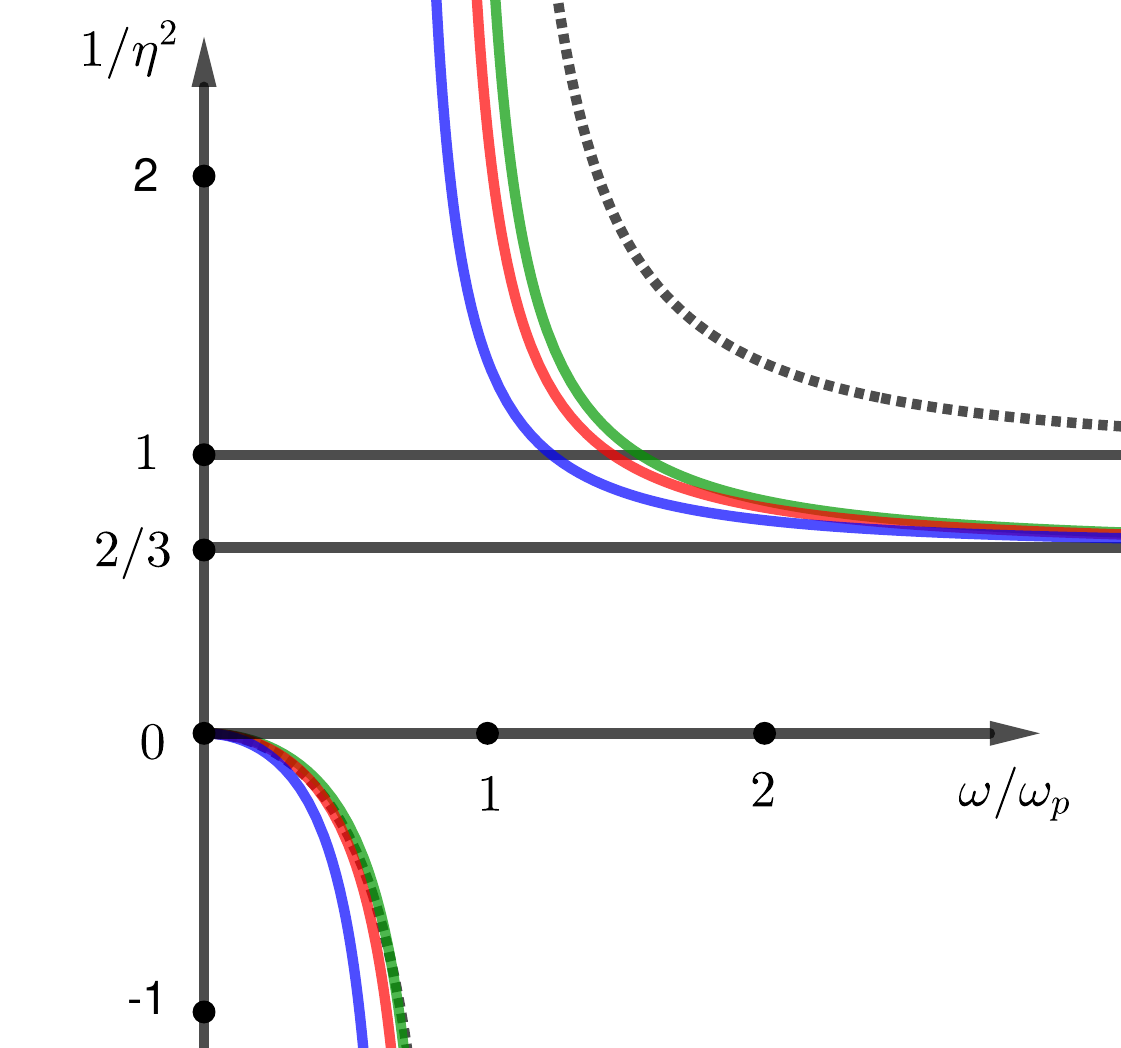}
\caption{Dispersion relations of the modified $O-$mode for particular cases of BI-type model with $p=1/2$ (green line), $p=3/4$ (red line) and $p \rightarrow \infty$ (blue line) in Eq. \eqref{modeO_BI-type}. The dotted line represents the usual $O-$mode. We assume $ \zeta =1$ and $c^2 B_0^2/\beta^2 = 1/2$. In the region $ 2/3 < 1/\eta^2 < 1 $, the wave travels slower than $c$.   }
\label{plot-modo-O}
\end{figure*}

For instance, the plot of $1/\eta^2$ versus $\omega/\omega_p$ with $ \zeta =1$ and $c^2 B_0^2/\beta^2 = 1/2$ is displayed in Fig. \ref{plot-modo-O}. We consider the  particular cases of $p=1/2$ (usual BI model), $p=3/4$ and $p \rightarrow \infty$ (Exponential model). Note that the propagating modes are divided into two regions with phase velocity $v_\phi > c$ and $v_\phi < c$. For the sake of comparison, we also exhibit the usual $O-$mode associated with Maxwell electrodynamics, where the wave travels only faster than $c$ (dotted line). Similar behaviour also occurs whenever $ \zeta > 0$, but for small values $  \zeta \, c^2 B_0^2/\beta^2 << 1$ the region with $v_\phi < c$ will decrease.

On the other hand, for $ \zeta < 0$ with the condition $| \zeta | < \beta^2/c^2 B_0^2 \,$ to guarantee a well-defined cut-off frequency, it is possible to show that $v_\phi > c$ throughout the propagation region.


The description of the modified $X-$mode can be performed in a similar way. 
Initially, by using the coefficients \eqref{c1_BI} and \eqref{d1_BI} in Eq. \eqref{X_mode_1}, we get 
\begin{equation} \label{modeX_BI-type}
\eta^2 = \frac{(\omega^2 - \widetilde{\omega}_L^2) (\omega^2 - \widetilde{\omega}_R^2) 
}{ \omega^2 (\omega^2 - \omega_c^2 - \widetilde{\omega}_p^2)}
\left[ \frac{  2p + c^2 B_0^2/\beta^2 }{ 2p + (2p - 1) c^2 B_0^2/\beta^2 } \right]
 \, , \end{equation}
where the modified plasma frequency is given by
\begin{equation} \label{wp_BI-type}
\widetilde{\omega}_p = \omega_p \left[ 1 + \frac{c^2 B_0^2}{2p \, \beta^2} \right]^{(1-p)/2}
\, , \end{equation}
and the modified cut-off frequencies $\widetilde{\omega}_L$ and $\widetilde{\omega}_R$ are defined in Eqs. \eqref{wl} and \eqref{wr}.

Notice that the dispersion relation \eqref{modeX_BI-type} does not depend on the parameter $ \zeta$. In addition,
we recall that the coefficients $c_1$ and $d_1$ modify the asymptotic behaviour for high frequency, which is obtained by the horizontal red line $(1 - d_1 B^2_0/c_1)$ in Fig. \ref{fig-modo-X}. For the BI-type model, this expression leads to
\begin{equation} \label{BI_type_hor_line}
1 - \frac{d_1 B_0^2}{c_1} = \frac{2p + (2p - 1) \, c^2 B_0^2/\beta^2}{2p + c^2 B_0^2/\beta^2} \, .
\end{equation}
Therefore, we conclude that the wave travels faster or slower than $c$ for high frequency, depending on whether $p>1$ or $0<p<1$, respectively.

Next, we proceed to describe the effects on the allowed and forbidden regions. From Eq. \eqref{wp_BI-type}, we have that $\widetilde{\omega}_p < \omega_p$  when $p>1$ and, consequently, the modified cut-off frequencies $\widetilde{\omega}_L$ and $\widetilde{\omega}_R$ also decrease in comparison with the standard results $\omega_L$ and $\omega_R$. However, 
one can promptly verify that  $\widetilde{\omega}_p$,  $\widetilde{\omega}_L$ and $\widetilde{\omega}_R$  increase for the parameter $0<p<1$.

Here, it is instructive to consider the weak field regime $\beta^2 >> c^2 B_0^2$. Bearing this in mind, for the allowed band $\widetilde{\omega}_L < \omega < \widetilde{\omega}_h$, we obtain
\begin{equation} \label{allowed_band}
\widetilde{\omega}_h - \widetilde{\omega}_L \approx \omega_h - \omega_L +
\frac{(p-1)}{2p} \, \omega_p^2 \, \frac{c^2 B_0^2}{\beta^2} 
\left[ \frac{1}{\sqrt{\omega_c^2 + 4 \omega_p^2}} - \frac{1}{2 \, \sqrt{\omega_c^2 +  \omega_p^2}} \right] 
\, . \end{equation}
From this result, one can easily see that the correction term is positive (negative) for the condition $p>1$ ($0<p<1$), leading to a bigger (smaller) allowed band. Similarly, for the forbidden band $\widetilde{\omega}_h < \omega < \widetilde{\omega}_R$, we find that
\begin{equation} \label{forbidden_band}
\widetilde{\omega}_R - \widetilde{\omega}_h \approx \omega_R - \omega_h +
\frac{(1-p)}{2p} \, \omega_p^2 \, \frac{c^2 B_0^2}{\beta^2} 
\left[ \frac{1}{\sqrt{\omega_c^2 + 4 \omega_p^2}} - \frac{1}{2 \, \sqrt{\omega_c^2 +  \omega_p^2}} \right] 
\, , \end{equation}
and arrive at the opposite previous conditions for the parameter $p$. These conclusions remain true by taking into account an expansion with arbitrary order.


\begin{figure*}[h]
\centering
\includegraphics[width=0.6\textwidth]{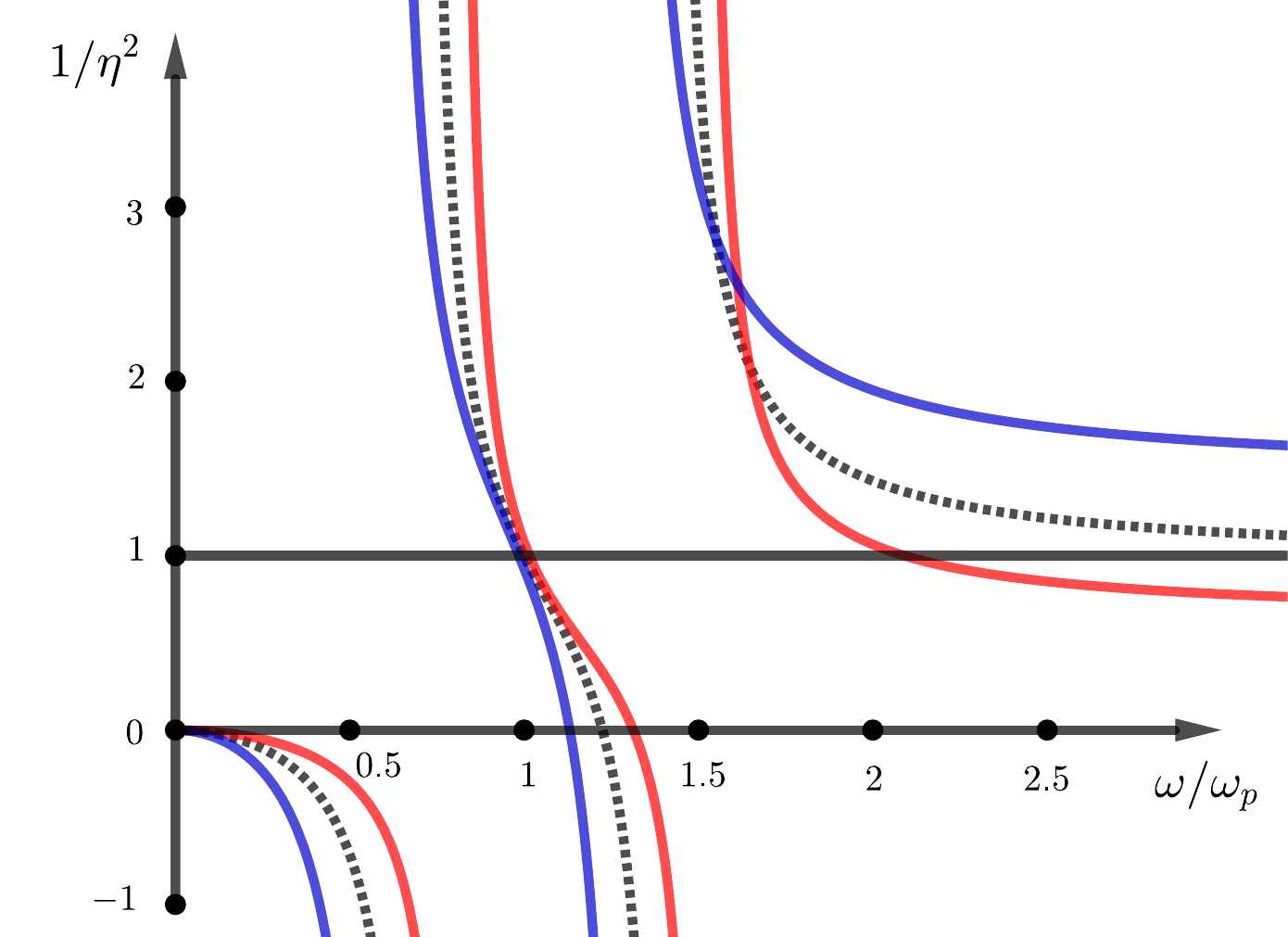}
\caption{Dispersion relations of the modified $X-$mode for BI-type model with $p=1/2$ (red line) and $p \rightarrow \infty$ (blue line) in Eq. \eqref{modeX_BI-type}. The dotted line represents the usual $X-$mode. We assume  $c^2 B_0^2/\beta^2 = 1/2$ and $\omega_c = \omega_p/\sqrt{2}$.   }
\label{plot-modo-X}
\end{figure*}

To illustrate the aforementioned results, the plot of $1/\eta^2$ versus $\omega/\omega_p$ with $c^2 B_0^2/\beta^2 = 1/2$ and $\omega_c = \omega_p/\sqrt{2}$ is exhibited in Fig. \ref{plot-modo-X}. We consider the  particular cases of $p=1/2$ (usual BI model) and $p \rightarrow \infty$ (Exponential model), described by the red and blue lines. By comparing with the usual $X-$mode from Maxwell electrodynamics (dotted line), we clearly see that the modified cut-off frequencies are shifted to the right (left) for  $p=1/2$ ($p \rightarrow \infty$). Furthermore, we would like to highlight the asymptotic behaviour for extremely large  $\omega$. In the case of $p=1/2$, the wave travels with $v_\phi < c$ because $1/\eta^2 \rightarrow 2/3$. As expected, for the usual $X-$mode, the phase velocity approaches to $c \, \;(1/\eta^2 \rightarrow 1)$. On the other hand, for $p \rightarrow \infty$ we have that $v_\phi > c$ and $1/\eta^2 \rightarrow 3/2$.


Finally, it is important to provide some physical estimates. According to the work \cite{Dehghani} and references therein, the $\beta-$parameter of BI-type electrodynamics is estimated in the range $10^{19}-10^{20} \, \textrm{V/m}$, or equivalently, $\beta/c \sim 10^{11}-10^{12} \, \textrm{T}$. Along this subsection, we have shown that $c^2 B_0^2/\beta^2$ plays a fundamental role in the description of the principal modes. Therefore, new contributions of BI-type models become more relevant with magnetic fields of the order $10^{10}-10^{12} \, \textrm{T}$ such that $c^2 B_0^2/\beta^2 \lesssim 1$. For example, this magnitude can be achieved in astrophysical environments of neutron stars, supernovae and gamma-ray bursts. However, in these scenarios, the analyses need to be complemented with a more detailed treatment by including an equation of state, as well as the relativistic effects and other contributions. Nevertheless, the relevance of the dimensionless parameter $c^2 B_0^2/\beta^2$ will certainly appear together with additional features.

 
\section{Relativistic and large amplitude effects} \label{sec_L_eff}

In this Section, we turn our attention to including relativistic and large amplitude effects. The circularly polarized waves possess some interesting properties, which allows us to solve exactly the system of fluid equations in the context of non-linear electrodynamics. In this case, one can easily incorporate a multicomponent plasma with electrons and ions motions.
Here, we adopt the approach in Eq. \eqref{L_approx}, by taking into account an expansion of non-linear electrodynamics. 
However, we focus on the leading-order corrections with quadratic terms in the invariants $\mathcal{F}$ and $\mathcal{G}$.
For the sake of completeness, let us also consider a parity-violating term proportional to $\mathcal{F} \mathcal{G}$.
Although it does not appear in most non-linear electrodynamics in the literature, we point out that the presence of dark matter candidates may generate an effective theory with such a term  (see, for instance, ref. \cite{dark_matter}). Therefore, we shall investigate the non-linear electrodynamics described by
\begin{equation} \label{eff_L_quad}
L_{nl} = \mathcal{F} + \frac{1}{B_c^2} \bigg[ \xi_1 \, \mathcal{F}^2 + \xi_2 \, \mathcal{G}^2
+ \xi_3 \, \mathcal{F} \mathcal{G} \bigg]
\, , \end{equation}
where we parametrized the non-linear contributions in terms of the critical magnetic field $B_c$, defined in Eq. \eqref{critical_E_B}, so that $\xi_1$, $\xi_2$ and $\xi_3$ are dimensionless parameters.

According to the definitions in Eqs. \eqref{def_D} and \eqref{def_H}, we find that
\begin{eqnarray}
{\bf D} & = &  \varepsilon_0 {\bf E} + \frac{\varepsilon_0}{B_c^2} \left\{ \left( \frac{ {\bf E} \cdot {\bf B} }{c} \right) 
\bigg[  \xi_3 \, {\bf E} + 2 \, \xi_2 \, c   {\bf B} \bigg] 
+  \left( \frac{ {\bf E}^2 }{c^2} - {\bf B}^2 \right) 
\left[ \xi_1 \, {\bf E} + \frac{1}{2} \, \xi_3 \, c  {\bf B} \right] \right\}
\label{D_quad} \, ,  \\
{\bf H} & = & \frac{ {\bf B} }{\mu_0} +  \frac{1}{\mu_0 B_c^2} \left\{ \left( \frac{ {\bf E} \cdot {\bf B} }{c} \right) 
\left[ \xi_3 \, {\bf B}  - 2 \, \xi_2 \, \frac{ {\bf E} }{ c} \right] 
+ \left( \frac{ {\bf E}^2 }{c^2} - {\bf B}^2 \right) 
\left[ \xi_1 \, {\bf B}  - \frac{1}{2} \, \xi_3 \, \frac{ {\bf E} }{ c} \right] \right\}
\label{H_quad} \, .
\end{eqnarray}

Before going ahead, it is opportune to discuss some properties of circularly polarized waves. 
Let us examine the RCP mode related to $ {\bf E} = E_0 \cos (\omega t - k z) \, \hat{x} + E_0 \sin (\omega t - k z) \, \hat{y} $. Even in the presence of the equilibrium magnetic field $ {\bf B}_0 = B_0 \hat{z} $, we have that 
${\bf E}^2 = E_0^2$ and ${\bf B}^2 = B_0^2 + k^2 E_0^2/ \omega^2$ are constant, as well as $ {\bf E} \cdot {\bf B} = 0 $.  With these simplifications, we now substitute the above constitutive relations in Eq. \eqref{eq_H}, which leads to the modified Amperè-Maxwell law
\begin{eqnarray} \label{eff_Ampere_Maxwell}
\nabla \times {\bf B}  
&+& \left( \frac{ {\bf E}^2 }{c^2} - {\bf B}^2 \right) \frac{1}{B_c^2} 
\left[ \xi_1 \left( \nabla \times {\bf B} - \frac{1}{c^2} \frac{\partial {\bf E}}{\partial t} \right) 
- \frac{1}{2} \frac{\xi_3}{c}  \left( \nabla \times {\bf E} + \frac{\partial {\bf B}}{\partial t} 
\right)  \right] = \nonumber \\
&=& \mu_0 \, {\bf j} + \frac{1}{c^2} \frac{\partial {\bf E}}{\partial t}
\, . \end{eqnarray}

Here, it should be noted that both parameters $\xi_2$ and $\xi_3$ do not contribute to this field equation. We call the attention that $\xi_3$ is multiplied by Faraday law, which remains unchanged in Eq. \eqref{Faraday_Gauss}. Keeping this in mind, one can rewrite Eq. \eqref{eff_Ampere_Maxwell} in the form
\begin{equation} \label{eff_Ampere_Maxwell_1}
\nabla \times {\bf B} = \mu_0 \, {\bf j}_{\textrm{eff}} + \frac{1}{c^2} \frac{\partial {\bf E}}{\partial t}  
\, , \end{equation}
where we defined the effective current density
\begin{equation} \label{eff_current}
{\bf j}_{\textrm{eff}} =  {\bf j} \, \left\{ 
1 - \xi_1 \left[ \left( \frac{E_0}{E_c} \right)^2 \left( \eta^2 - 1 \right) 
+ \left( \frac{B_0}{B_c} \right)^2 \right] \right\}^{-1} \, .
\end{equation}

After these manipulations, we are now ready to apply a similar methodology of ref. \cite{Stenflo_1976} for Maxwell electrodynamics, but using the aforementioned effective current density instead of the usual one. According to ref. \cite{Borovsky}, the relativistic fluid equations of cold plasma are given by
\begin{eqnarray}
\frac{\partial n_s}{\partial t} + \nabla \cdot \left( n_s \, {\bf u}_s \right) &=& 0 \, , \label{eq_n_rel} \\
\frac{\partial{\bf p}_s}{\partial t} + \left( {\bf u}_s \cdot \nabla \right) {\bf p}_s &=& 
q_s \, ({\bf E} + {\bf u}_s \times {\bf B}) \, , \label{eq_F_L_rel} 
\end{eqnarray}
where $s$ is the index of particle species (electrons and ions) and $ {\bf p}_s = \gamma_s \, m_s \, {\bf u}_s$ denotes the relativistic momentum with the Lorentz factor  $\gamma_s = 1/ \sqrt{ 1 - {\bf u}_s^2/c^2} $. In addition, $m_s$ and $q_s$ correspond to the mass and charge of each species, respectively.

By adopting the standard notation 
$E_{\pm} = E_x \pm i E_y$, $B_{\pm} = B_x \pm i B_y$ and $u_{\pm} = u_x \pm i u_y$, one can promptly show that 
the fluid equations \eqref{eq_n_rel} and \eqref{eq_F_L_rel} are satisfied with $n_s = n_{0s}$ being constant, $u_z=0$ and
\begin{equation} \label{sol_fluid}
E_{\pm} = E_0 \exp( \pm i \omega t \mp i k z) \, \; , \, \; 
B_{\pm} = \pm i \frac{k}{\omega} \, E_{\pm} \, \; , \, \; 
u_{\pm} = \mp \frac{i q_s}{\gamma_s m_s (\omega + \omega_{cs})} E_{\pm} \; \, , 
\end{equation}
where $\omega_{cs} = q_s B_0/\gamma_s m_s$ is the relativistic cyclotron frequency. Furthermore, the Lorentz factor obeys the algebraic equation
\begin{equation} \label{gamma_relation}
\gamma^2_s = 1 + \frac{q_s^2 E_0^2}{m_s^2 c^2 (\omega + \omega_{cs})^2 }  \, \; ,
\end{equation}
or equivalently,
\begin{equation} \label{gamma_relation_1}
\omega^2 \gamma^4_s + 2 \omega \left( \frac{q_s B_0}{m_s} \right) \gamma^3_s + 
\left[ \left( \frac{q_s B_0}{m_s} \right)^2 - \left( \frac{q_s E_0}{m_s c} \right)^2 - \omega^2  \right] \gamma^2_s
- 2 \omega \left( \frac{q_s B_0}{m_s} \right) \gamma_s - \left( \frac{q_s B_0}{m_s} \right)^2 = 0 \, .
\end{equation}
In this case, the non-trivial contribution of the Lorentz factor appears due to the large amplitude effect.
Observe that, for small amplitude wave, we have that $\gamma_s \approx 1$.

At this stage, we insert these results in Eq. \eqref{eff_Ampere_Maxwell_1} with the current density $ {\bf j} = \sum_s n_s q_s {\bf u}_s $, which yields the following dispersion relation

\begin{equation} \label{eff_DR}
\eta^2 - 1 = - \sum_s \frac{\omega_{ps}^2}{\omega (\omega + \omega_{cs})} + 
\xi_1 \left[ \left( \frac{E_0}{E_c} \right)^2 \left( \eta^2 - 1 \right) 
+ \left( \frac{B_0}{B_c} \right)^2 \right] \left( \eta^2 - 1 \right) \, .
\end{equation}
where $\omega_{ps} = \sqrt{n_{0s} \, q_s^2/ \gamma_s m_s \varepsilon_0}$ denotes the relativistic plasma frequency. 

Similarly, one can get the LCP mode through the replacement of $\omega_{cs} \rightarrow - \omega_{cs}$ in the previous expression. In addition, we mention that, for the particular case of Euler-Kockel electrodynamics $(\xi_1 = 2 \alpha/45\pi)$ 
with one-component electron fluid moving in a fixed ionic background, one can easily recover the result 
in Eq. \eqref{RCP_EK} by disregarding the large amplitude contribution.

The general solution of the dispersion relation \eqref{eff_DR} is quite involved, but we can obtain interesting results even in the unmagnetized case $(B_0=0)$. From now on, let us consider this assumption. Therefore, using $\omega_{cs} = 0$ in Eq. \eqref{gamma_relation}, we get the relativistic Lorentz factor $ \gamma_s = \sqrt{1 + ( q_s E_0/m_s c \, \omega)^2} $.
Without loss of generality, we also assume $\xi_1 << 1$ such that the dispersion relation can be approximated by 
\begin{equation} \label{eff_DR_unmag}
\eta^2 - 1 \approx - \sum_s \frac{\omega_{ps}^2}{\omega^2} 
+ \xi_1 \left( \frac{E_0}{E_c} \right)^2 \left[  \sum_s \frac{\omega_{ps}^2}{\omega^2} \right]^2 
\, . \end{equation}
Next, we define the effective plasma frequency
\begin{equation} \label{eff_w_unmag}
\omega_{p,0} = \sqrt{\sum_s \omega_{ps}^2}
\,  \end{equation}
and the refractive index in the absence of non-linear electrodynamics $(\xi_1 =0)$,
\begin{equation} \label{eff_eta_unmag}
\eta_{p,0} = \sqrt{ 1 - \frac{\omega_{p,0}^2}{\omega^2}  }
\, . \end{equation}

With these definitions, Eq. \eqref{eff_DR_unmag} can be recast as
\begin{equation} \label{eff_DR_unmag_1}
\eta^2  \approx \eta_{p,0}^2
+ \xi_1 \left( \frac{E_0}{E_c} \right)^2 \left[  1 - \eta_{p,0}^2 \right]^2 
\, . \end{equation}

Notice that, for small plasma densities $\eta_{p,0} \rightarrow 1$, the non-linear contributions disappear. On the other hand, 
when the frequency of the propagating wave is close to the effective plasma frequency $\omega \rightarrow \omega_{p,0} \,$, or equivalently $\eta_{p,0} \rightarrow 0 $, we observe that the refractive index is mainly fixed by the correction of non-linear electrodynamics, namely, $ \eta \rightarrow \sqrt{\xi_1} \, (E_0/E_c) $. This non-vanishing behaviour was initially obtained in the work of ref. \cite{Piazza_POP_2007}, where the authors considered the particular case of Euler-Kockel electrodynamics
and assigned it to the enhancement of vacuum polarization effects. Here, we generalize this result for the non-linear effective model in Eq. \eqref{eff_L_quad}, which encompasses the leading-order corrections of most examples in the literature. 
Thus, we conclude that only the contributions related to the parameter $\xi_1$ will modify the dispersion relation and, consequently, the refractive index.  

Finally, it is important to highlight that the above non-vanishing behaviour of the refractive index always occurs when $\xi_1>0$. This condition is automatically satisfied by Euler-Kockel electrodynamics. Moreover, for the effective Born-Infeld-type model described in Eq. \eqref{L_BI_approx}, we recognize that $\xi_1 = (E_c/\beta)^2 \, (1-p)/2p $ and only the particular cases with $0<p<1$  will exhibit an analogous behaviour, which includes the usual Born-Infeld electrodynamics $(p=1/2)$. However, the exponential electrodynamics $(p \rightarrow \infty)$ does not obey the previous condition.

\section{Concluding Comments and Perspectives} \label{sec_concl}

In this contribution, we have investigated some modified plasma waves in the context of non-linear electrodynamics. 
In Secs. \ref{sec_NL_ED} and \ref{sec_applic}, we have considered a magnetized plasma with electrons and a uniform ionic background within a cold fluid model. In addition, we disregarded the collisional, relativistic and high amplitude effects. The dispersion relations were obtained in terms of the coefficients $c_1$, $d_1$ and $d_2$, which are determined by specifying the electrodynamics under consideration, as described in Eqs. \eqref{coeff_c} and \eqref{coeff_d}.  We have assumed parity symmetry, which implies $d_3=0$. In what follows, our results are summarized.

For electrostatic waves, we arrived at the modified Trivelpiece-Gould dispersion relation by substituting the plasma frequency
$\omega_p \rightarrow \bar{\omega}_p \,$, defined in Eq. \eqref{w_bar_p}. To guarantee the stability of the corresponding modes, one has the following constraint $c_1 + d_2 B^2_0 \cos^2 \theta > 0$,  where $\theta$ denotes the angle between the wave propagation direction and magnetic field ${\bf B}_0$. We have also found a generalized Appleton-Hartree equation and investigated its principal modes. For circularly polarized waves (RCP and LCP modes), the standard analysis applies with the replacement $\omega_p \rightarrow \widetilde{\omega}_p = \omega_p/\sqrt{c_1}$. Moreover, for the modified ordinary $(O)$ mode, Eq. \eqref{O_mode}, we recognized a cut-off at 
$\omega = \omega_p/\sqrt{c_1 + d_2 B_0^2}$ and obtained the asymptotic behaviour for high frequency, where $1/\eta^2$ approaches to $1/(1 + d_2 B_0^2/c_1)$, with $\eta$ being the refractive index. The qualitative description is exhibited in Fig. \ref{fig-modo-O}. Similarly, for the modified extraordinary $(X)$ mode, we determined the resonance and cut-off frequencies with the same change in the plasma frequency $(\omega_p \rightarrow \widetilde{\omega}_p)$. Keeping this in mind, a resonance occurs at the modified upper-hybrid frequency $\widetilde{\omega}_h $ and the cut-off frequencies are located at $\widetilde{\omega}_L $ and 
$\widetilde{\omega}_R $, defined in Eqs. \eqref{wl} and \eqref{wr}. Again, we observed a modification in the asymptotic behaviour for high frequency, given by $1/\eta^2 \rightarrow (1 - d_1 B_0^2/c_1)$. The qualitative description is illustrated in Fig. \ref{fig-modo-X}. As expected, the usual dispersion relations from Maxwell electrodynamics are recovered when $c_1 =1$ and $d_1 = d_2 = 0$.

To check the consistency of our results and clarify the methodology, we have considered the well-known Euler-Kockel electrodynamics and showed that the corresponding dispersion relations are in agreement with the literature. Furthermore, we also investigated the Born-Infeld-type electrodynamics, which encompasses a set of models defined in the parameter space $(p,  \zeta, \beta)$. For each dispersion relation, we found some constraints involving these parameters. Below, the main results are pointed out.

In the case of electrostatic waves, the dispersion relation is well-defined for $ \zeta \geq 0$. It is also possible to have $ \zeta < 0$ with the constraint $| \zeta | \cos^2 \theta < \beta^2/c^2 B_0^2$. Interestingly enough, in the weak field regime, 
the new effects cancel out when the parameters sa\-tisfy  $(1-p)/2p =  \zeta \cos^2 \theta$. As already mentioned, for circularly polarized waves, the new contribution is included into the modified plasma frequency $\widetilde{\omega}_p$. By considering weak fields, we obtained a similar expression of Euler-Kockel electrodynamics and showed that the correspondent contribution may assume positive or negative values depending on whether $p>1$ or $0<p<1$. Next, for the modified ordinary mode, an interesting situation occurs whenever $ \zeta >0$, namely, the propagation region allows phase velocity smaller than $c$, as described in Fig. 
\ref{plot-modo-O}. We remember that it does not happen for the usual $O-$mode. At last but not least, we analysed the modified extraordinary mode in which the parameter $p$ plays a fundamental role. First, the asymptotic behaviour for high frequency is altered, such that the wave travels faster or slower than $c$ in accordance with $p>1$ and $0<p<1$, respectively. We also emphasized that these conditions provide different effects on the allowed and forbidden regions, which may be smaller or greater than the standard results. The particular cases with $p=1/2$ and $p \rightarrow \infty$ are displayed in Fig. \ref{plot-modo-X}.

Next, in Sec. \ref{sec_L_eff}, we have considered the inclusion of relativistic and large amplitude effects in circularly polarized waves. For non-linear electrodynamics up to quadratic terms in the invariants $\mathcal{F}$ and $\mathcal{G}$, we have obtained an exact solution of the system, given by the dispersion relation \eqref{eff_DR}. In the unmagnetized situation, when the wave propagation frequency approaches the effective plasma frequency, we have shown that the refractive index is mainly established by the non-linear corrections related to $\mathcal{F}^2$. We also found the condition $\xi_1 > 0$ to guarantee this behaviour, which is satisfied only in specific models, such as Born-Infeld-type with $0<p<1$ and Euler-Kockel electrodynamics.

To conclude, we would like to point out some perspectives. For instance, let us focus on the results and assumptions of Sec. \ref{sec_L_eff}. Firstly, it should be mentioned that, for the unmagnetized case with Euler-Kockel electrodynamics discussed in ref. \cite{Piazza_POP_2007}, the authors showed that the inclusion of collisional contributions does not change the results on the enhancement of vacuum polarization effects. Therefore, it remains to be examined in the context of non-linear electrodynamics described in Eq. \eqref{eff_L_quad}. Moreover, in the magnetized situation, we have obtained a non-trivial algebraic equation for the relativistic Lorentz factor, Eq. \eqref{gamma_relation_1}, which certainly deserves further investigation.  New plasma modes or modifications to the usual ones may appear due to non-linear corrections. In addition, we also expect that stability criteria will be modified. We hope that the results presented here can be useful to pursue some investigations in these directions and to find applications in experiments involving strong laser fields, where the approximation of cold plasma is well justified. We expect to report on these issues elsewhere.


{\bf Acknowledgments:} the authors acknowledge the support by {\it Conselho Nacional de Desenvolvimento Científico e Tecnológico} (CNPq). In particular, L. P. R. Ospedal is grateful for a post-doctoral fellowship under grant 166386/2020-0, when a part of this work was carried out. 

{\bf Data Availability Statement:} the data that support the findings of this study are available from the corresponding author upon reasonable request.


\end{document}